\documentclass[a4paper,11pt]{revtex4-1}
\usepackage{amsmath}
\usepackage{amsfonts}
\usepackage{amssymb}
\usepackage{graphicx}
\usepackage{here}

\begin{document}

\title{Hierarchical Prisoner's Dilemma in Hierarchical Public-Goods Game}
\author{Yuma Fujimoto$^{1, \dag}$, Takahiro Sagawa$^{2}$, and Kunihiko Kaneko$^{1}$}
\affiliation{$^{1}$Department of Basic Science, The University of Tokyo, 3-8-1, Komaba, Meguro-ku, Tokyo 153-8902, Japan\\
$^{2}$Department of Applied Physics, The University of Tokyo, 7-3-1, Hongo, Bunkyo-ku, Tokyo 113-8654, Japan}
\affiliation{$^{\dag}${\rm yfujimoto@complex.c.u-tokyo.ac.jp}}

\begin{abstract}
{The dilemma in cooperation is one of the major concerns in game theory. In a public-goods game, each individual pays a cost for cooperation, or to prevent defection, and receives a reward from the collected cost in a group. Thus, defection is beneficial for each individual, while cooperation is beneficial for the group. Now, groups (say, countries) consisting of individual players also play games. To study such a multi-level game, we introduce a hierarchical public-goods (HPG) game in which two groups compete for finite resources by utilizing costs collected from individuals in each group. Analyzing this HPG game, we found a hierarchical prisoner's dilemma, in which groups choose the defection policy (say, armaments) as a Nash strategy to optimize each group's benefit, while cooperation optimizes the total benefit. On the other hand, for each individual within a group, refusing to pay the cost (say, tax) is a Nash strategy, which turns to be a cooperation policy for the group, thus leading to a hierarchical dilemma. Here, the reward received by one group increases with the population, as does the collected cost. In spite of this, we find that there exists an optimal group size that maximizes its payoff. Furthermore, when the population asymmetry between two groups is large, a smaller group will choose a cooperation policy (say, disarmament) to avoid excessive response from the larger group, which leads to the resolution of the prisoner's dilemma between the groups. The relevance of the HPG game to policy selection in society and the optimal size in human or animal groups are discussed accordingly.}
\end{abstract}

\maketitle

\section*{Introduction}
Hierarchical structures are ubiquitous in society. For example, a human society or a country consists of people, while the world or a higher group of societies consists of countries or lower-level groups. Such a hierarchy also exists in some animal societies, where herds in a region interact with each other. Within a group or country, individuals may cooperate or defect, whereas a group (country) chooses some policy to interact with other groups. Thus, interplay between intra-group strategies of individuals and inter-group policy is important in understanding the social structure of cooperation.

In considering cooperation among individuals within a group, the public goods (PG) game is commonly adopted\cite{Frank1998, Levin2014, Wakano2009}. In the PG game, each individual has to pay a certain cost for the goods in the society, while not paying the cost will be advantageous for the individual. How cooperation in a society is achieved has been extensively studied in the PG game\cite{Axelrod1981, Hauert2002, Santos2008}. For example, each person in a country is asked to pay a tax for the country, with which each can get equal welfare as payoffs from the country, depending on the total taxes collected. Here the total payoff of the country is maximized when all individuals cooperate in paying the required taxes. This achieves the Pareto optimum in an intra-group game. On the other hand, each individual can choose to free-ride, i.e., tries to receive the payoff without paying the tax, for his or her own benefit. Indeed, the Nash equilibrium\cite{Nash1951}, or the optimal strategy, for an individual is achieved when all people free-ride which, however, results in smaller payoffs. This situation is common to the standard prisoner's dilemma\cite{Tucker1950, Hardin1968}.

To study cooperation in a hierarchical society, however, we need to consider inter-group games also, where groups compete for resources with each other. As an example, we consider a simple allegorical situation in which each country struggles for finite resources using arms, for which taxes are collected from the people. Each country faces a strategic choice: either compete for resources participating in arms race or cooperate and divide the resources without suffering the losses from war. However, each individual has to choose whether to cooperate by paying tax or evade it. In this sense, the game in question is hierarchical in nature. A defection at the individual level leads to cooperation at the group level, whereas cooperation at the individual level in paying tax leads to a struggle at the group level in arms race. In terms of game theory, the former implies a Pareto equilibrium between groups, whereas the latter leads to a Nash equilibrium for each group. Disagreement between the two strategies implies a prisoner's dilemma, which exists across levels as well. It is important to formulate this hierarchical game with a prisoner's dilemma across levels.

In order to reveal the fundamental features of hierarchical societies, we introduce and analyze a new prototype for a multi-level game, which we refer to as the hierarchical public-goods (HPG) game. In the HPG game, groups compete for resources, and individuals simultaneously play the PG game within each group. Here, note that the ability to obtain resources generally depends on the group size, (i.e., population size of the group). Hence, this group size effect is introduced as the dependence of the individual payoff on the group size.

We analyze the optimal choice of strategy both at the individual and group levels, and find a novel type of dilemma intrinsic to the hierarchical game, which we call the hierarchical prisoner's dilemma. We show that a hierarchical prisoner's dilemma emerges when the two group sizes are not largely different. However, with a large group size difference, defection (i.e., refusal to pay the cost) is favored at the individual level in a relatively smaller group, and the hierarchical dilemma is avoided.

There have been previous studies on multi-level games in which groups compete with each other following individual players' actions\cite{Rapoport1987, Hausken2000, Hausken2004}. Although these studies have discussed context dependence of cooperation and defection on levels, they have not introduced an action strategy at the group level in their games. However, Traulsen and Nowak\cite{Traulsen2006} studied multi-level selection of groups consisting of individuals that either cooperate or defect. Groups are selected depending on the fitness computed from the action of individuals, where a defector (parasite) gets an advantage at the individual level, but a group dominated by such individuals has lower fitness and is eliminated. With this multi-level selection, they obtained a condition for cooperative individuals to be fixed. In their study, however, the groups do not play a game with each other, and their fitness is just given by the actions of individuals within the group. In a hierarchical society, in contrast, each group (e.g., a country) plays an inter-group game, and chooses an action depending on its own policy. Our model, thus, provides a general framework for multi-layer competition.

\section*{Model of the HPG game}
We now introduce a model for the HPG game, in which two groups compete with each other for a restricted amount of resource. (Throughout the present paper, the number of groups is set to two, but extension to more groups is straightforward.) An individual belongs to one of the groups and pays a cost to his or her own group. Each group utilizes all the costs collected from the individuals in the group to win the competition with the other groups. We assume that the competitive capacity of a group increases with the number of individuals, in addition to the summed cost, considering the collective effect of the population. To be specific, the capacity is assumed to be proportional to both the $\alpha$-th power of the population and the summed cost. Here, $\alpha$ is a positive number that characterizes the efficiency of utilizing the cost for the competition: the larger $\alpha$ is, the more the advantage of the larger group.

This $\alpha$-th power of the population follows, for example, Lanchester's law\cite{Lanchester1916}, in which the strength of military forces increases with the $\alpha$-th power of the population.
For example, $\alpha=1$ holds in a battle on a narrow bridge, and $\alpha=2$ in a wide field. Furthermore, this type of law is also applicable to competitions in biological groups\cite{McGlynn2000, Wilson2002, Shelley2004}. McGlynn et al.\cite{McGlynn2000} carried out an experiment of ants to compete for foods in pallets and found that a larger (smaller) group has more advantage when the entrance of pallets is wider (narrower). In this case, the entrance size could correspond to the exponent $\alpha$.

We now formulate the HPG explicitly. Consider a situation in which $N$ individuals are divided into two groups with populations $(N_1,N_2)$ and $N_1+N_2=N$. Without loss of generality, group 1 is assumed to have a larger population, i.e., $N_1 \ge N_2$, throughout the paper. The payoff of individual $j \in \{ 1, \cdots , N_i \}$ in the group $i \in \{ 1,2 \}$ is defined as
\begin{equation}
u_{ij} = \frac{1}{N_i} \frac{X_iN_i^{\alpha}}{X_1N_1^{\alpha}+X_2N_2^{\alpha}} M - x_{ij}.
\label{payoff}
\end{equation}
Here, $x_{ij}$ $(\ge 0)$ is the cost paid by individual $j$ in group $i$, while $X_i := \sum_j x_{ij}$ is the sum of individuals' costs in the group $i$. $M$ is the total amount of the resource, which is divided into the two groups according to the capacity of each group, which is proportional to $X_iN_i^{\alpha}$. The first term in $u_{ij}$ gives the reward, that is, the resource distributed among the individuals according to the group's total cost, while the second term gives the cost paid by each individual. The payoff function is defined as the reward reduced by the cost paid. The payoff function Eq.~1 in HPG is schematically shown in Fig.~1.

\begin{figure}
\begin{center}
\includegraphics[width=0.5\textwidth]{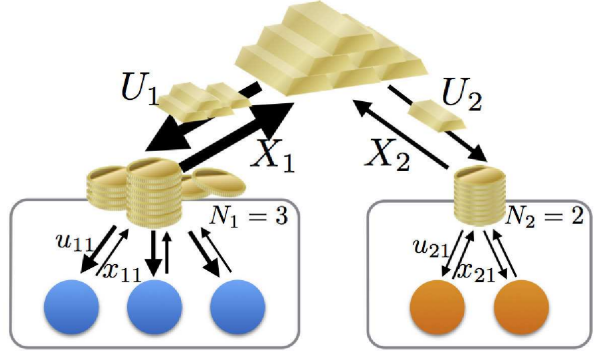}
\caption{Schematic diagram of the HPG game.
Individuals $(i,j)$ belong to one of the two groups $i \in \{ 1,2 \}$, and $j \in \{1, \cdots, N_i \}$, where $N_i$ is group $i$'s size.
The power of each group to obtain a resource is given by $X_iN_i^{\alpha}$, where $X_i$ is the accumulated cost paid by the individuals in group $i$.
}
\end{center}
\end{figure}

Now the average payoff in all group games, written as $u_{\mathrm{ave}}$, is given by
\begin{align}
	u_{\mathrm{ave}} := \frac{1}{N} \sum_{ij} u_{ij} = \frac{M}{N} - \frac{1}{N} \sum_{ij} x_{ij} .
\end{align}
From Eq.~2, when $x_{ij} = 0$ holds for any $i,j$, the average payoff $u_{\mathrm{ave}}$ takes the maximum value. In other words, when all individuals do not pay any cost, the groups do not compete with each other, and the total average payoff is maximal. This implies that $x_{ij}=0$ for any $i,j$ is the Pareto optimum.

\section*{Game between groups}
Before considering the role of individuals in the HPG game, we study a characteristic feature of the above inter-group game. Here, groups 1 and 2 compete with each other for a restricted amount of resource $M$, by paying costs $X_1$ and $X_2$. Using the HPG model (Eq.~1), the payoff function of group $i$ is defined as
\begin{align}
	& U_i = \frac{X_iN_i^{\alpha}}{X_1N_1^{\alpha}+X_2N_2^{\alpha}} M - X_i .
\end{align}

We now consider the Nash equilibrium, where each player is a group, not an individual. The cost of the group Nash equilibrium, denoted as $(X_1^{\mathrm{GN}}, X_2^{\mathrm{GN}})$, satisfies
\begin{align}
	\nonumber
	& \left. \frac{\partial U_i}{\partial X_i} \right|_{\{ X_i = X_i^{\mathrm{GN}}| i \in \{ 1,2 \} \}} = 0
\end{align}
for any $i$.
From this, $(X_1^{\mathrm{GN}}, X_2^{\mathrm{GN}})$ is given by
\begin{align}
	\nonumber
	& X_i^{\mathrm{GN}} = \frac{N_1^{\alpha}N_2^{\alpha}}{(N_1^{\alpha}+N_2^{\alpha})^2} M .
\end{align}
Then, each group's payoff in the Nash equilibrium, denoted by $(U_1^{\mathrm{GN}}, U_2^{\mathrm{GN}})$, is given by
\begin{align}
	\nonumber
	& U_i^{\mathrm{GN}} = \frac{N_i^{2\alpha}}{(N_1^{\alpha}+N_2^{\alpha})^2} M .
\end{align}
In the Nash equilibrium, both the groups need to make substantial payments because if one group pays less, more resources are taken by the other. However, if both the groups agree not to pay the cost for the struggle, i.e., succeed in cooperation, all the resources are distributed without any costs. In this case, they can achieve the Pareto optimum, whose cost $(X_1^{\mathrm{GP}}, X_2^{\mathrm{GP}})$ is given by
\begin{align}
	\nonumber
	& X_i^{\mathrm{GP}} = 0 .
\end{align}
In this case, resource allocation to each group is indefinite in Eq.~3.
However, if we take the limit $X_i\rightarrow 0$, keeping $X_1=X_2$, then each group's payoff in the (group) Pareto optimum $(U_1^{\mathrm{GP}}, U_2^{\mathrm{GP}})$ is given by
\begin{align}
	\nonumber
	& U_i^{\mathrm{GP}} = \frac{N_i^{\alpha}}{N_1^{\alpha}+N_2^{\alpha}} M .
\end{align}
Since both $X_i^{\mathrm{GN}}>X_i^{\mathrm{GP}}$ and $U_i^{\mathrm{GN}}<U_i^{\mathrm{GP}}$ hold, there exists a prisoner's dilemma, i.e., the payoff of the Nash equilibrium is smaller than the Pareto optimum.

As an example of inter-group prisoner's dilemma, consider the case in which each country attempts to pay more costs for arms to compete with other countries. This leads to an arms race, resulting in paying more army expenditure as in a Nash equilibrium, while they can obtain resources without costs as in a Pareto equilibrium, if the countries somehow disagree for arms race.

\section*{Game between individuals}
We next focus on the cost and payoff of each individual. First, we consider the {\bf intra}-group game of individuals only within a single group (i.e., group 1, the larger group), whereas the total cost in the other group $X_2$ is fixed. Then, we obtain the cost of individuals in group 1 who satisfy the Nash versus Pareto strategies, which correspond to the group 1 policies of defection and cooperation, respectively.

First, we consider the Nash equilibrium at the individual level, in which each individual in group 1 pursues his or her own benefit. As will be shown, this strategy is to defect for the benefit of the group; for simplicity, it is denoted as Defection (D). Then, individual $j$ in group 1 determines his or her own cost as a function of $X_2$, denoted as $x_{1j}=x_{1j(\mathrm{D})}(X_2)$, to maximize one'€™s own payoff $u_{1j}$, whose condition is given by
\begin{align}
	\nonumber
	& \left. \frac{\partial u_{1j}}{\partial x_{1j}} \right|_{\{ x_{1j}=x_{1j(\mathrm{D})}(X_2) | j \in \{ 1, \cdots, N_1 \} \}} = 0 \quad (\forall j) \\
	\nonumber
	& \Leftrightarrow X_{1(\mathrm{D})}(X_2) = \sqrt{\frac{M}{N_1} \left( \frac{N_2}{N_1} \right)^{\alpha}X_2} - \left( \frac{N_2}{N_1} \right)^{\alpha}X_2 .
\end{align}
Here, $X_{1(\mathrm{D})}(X_2) := \sum_j x_{1j(\mathrm{D})}(X_2)$ is the total cost in group 1 when all individuals defect.
$X_{1(\mathrm{D})}(X_2)$ represents the total cost in the individual Nash equilibrium as a function of the fixed cost of the other group.
Then, the total payoff in group 1, denoted as $U_{1(\mathrm{D})}(X_2)$, is given by
\begin{align}
	\nonumber
	& U_{1(\mathrm{D})}(X_2) = \frac{X_{1(\mathrm{D})}(X_2)N_1^{\alpha}}{X_{1(\mathrm{D})}(X_2)N_1^{\alpha}+X_2N_2^{\alpha}} M - X_{1(\mathrm{D})}(X_2) .
\end{align}

Second, we consider the Pareto equilibrium in each individual's game in group 1. In this case, each individual pays a cost to maximize the benefit of group 1, so that it is simply denoted as Cooperative (C). Now, to maximize the group's payoff $U_1$, the cost of individual $j$ in group 1, $x_{1j}=x_{1j(\mathrm{C})}(X_2)$, has to satisfy the condition
\begin{align}
	\nonumber
	& \left. \frac{\partial U_1}{\partial X_1} \right|_{\{ x_{1j}=x_{1j(\mathrm{C})}(X_2) | j \in \{ 1, \cdots, N_1 \} \}} = 0 \quad (\forall j) \\
	\nonumber
	& \Leftrightarrow X_{1(\mathrm{C})}(X_2) = \sqrt{M \left( \frac{N_2}{N_1} \right)^{\alpha}X_2} - \left( \frac{N_2}{N_1} \right)^{\alpha}X_2,
\end{align}
where $X_{1(\mathrm{C})}(X_2) := \sum_j x_{1j(\mathrm{C})}(X_2)$ is the total cost in group 1 when all individuals cooperate.
$X_{1(\mathrm{C})}(X_2)$ represents the total cost in the individual Pareto optimum. Then, the total payoff in group 1, denoted as $U_{1(\mathrm{C})}(X_2)$, is given by
\begin{align}
	\nonumber
	& U_{1(\mathrm{C})}(X_2) = \frac{X_{1(\mathrm{C})}(X_2)N_1^{\alpha}}{X_{1(\mathrm{C})}(X_2)N_1^{\alpha}+X_2N_2^{\alpha}} M - X_{1(\mathrm{C})}(X_2).
\end{align}

\begin{figure}
\begin{center}
\includegraphics[width=0.5\textwidth]{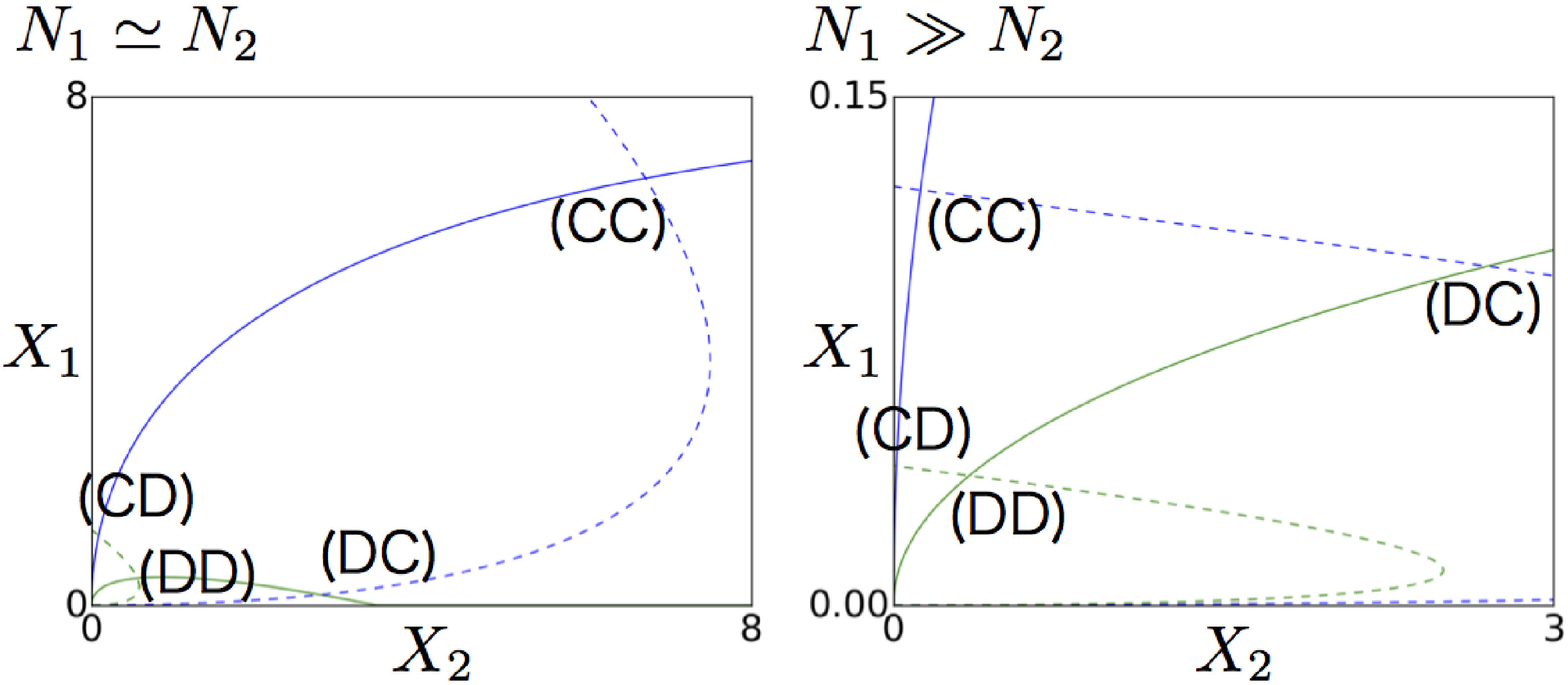}
\caption{Each group's total cost $(X_1, X_2)$ for four policies, $(\mathrm{CC}), (\mathrm{CD}), (\mathrm{DC}), and (\mathrm{DD})$.
The blue (green) solid line indicates that $X_1$ responds to $X_2$ when group 1 chooses C (D), while the blue (green) broken line indicates that $X_2$ responds to $X_1$ when group 2 chooses C (D).
The intersections of each solid and broken line indicate the equilibrium points. The left (right) figure is for $(N_1,N_2)=(17,13)$ ($(27,3)$).
The payoff of each group is determined by each equilibrium point.
}
\end{center}
\end{figure}

From a comparison of the above two cases, $X_{1(\mathrm{C})}(X_2) > X_{1(\mathrm{D})}(X_2)$ and $U_{1(\mathrm{C})}(X_2) > U_{1(\mathrm{D})}(X_2)$ hold. Thus, there exists a prisoner's dilemma in the individual game inside a group (see Supporting Information): When the individuals in group 1 cooperate, the group collects more costs from its members and receive a higher payoff, but an attempt to minimize its own cost leads to the less beneficial Nash equilibrium. Individuals do not pay sufficient costs to compete with the other group if they defect, while the payment is sufficient for cooperation.

The same argument applies to group 2 to determine $X_{2(\mathrm{D})}(X_1)$, $U_{2(\mathrm{D})}(X_1)$, $X_{2(\mathrm{C})}(X_1)$, $U_{2(\mathrm{C})}(X_1)$.
In addition, $X_{2(\mathrm{C})}(X_1) > X_{2(\mathrm{D})}(X_1)$ and $U_{2(\mathrm{C})}(X_1) > U_{2(\mathrm{D})}(X_1)$ also hold.

Third, we consider a situation in which each group can choose either cooperation or defection for individuals. In this case, ``C" and ``D" are regarded as the ``policies" of each group. Now, the two groups have totally four choices, $(\mathrm{YZ})$ with $\mathrm{Y} \in \{ \mathrm{C,D} \}$ and $\mathrm{Z} \in \{ \mathrm{C,D} \}$, where $\mathrm{Y}$ ($\mathrm{Z}$) indicates the policy in group 1 (2). Then, the equilibrium total cost in each group, denoted as $(X_{1(\mathrm{YZ})}, X_{2(\mathrm{YZ})})$, satisfies 
\begin{align}
	\nonumber
	& X_{1(\mathrm{YZ})} = X_{1(\mathrm{Y})}(X_{2(\mathrm{YZ})}) , \\
	\nonumber
	& X_{2(\mathrm{YZ})} = X_{2(\mathrm{Z})}(X_{1(\mathrm{YZ})}) .
\end{align}
In this way, each equilibrium point $(X_{1(\mathrm{YZ})}, X_{2(\mathrm{YZ})})$ is determined from the balance of costs between the two groups (see Supporting Information). In other words, it is given as a cross point of two functions, $X_{1(\mathrm{Y})}(X_{2})$ and $X_{2(\mathrm{Z})}(X_{1})$, in the $X_1$-$X_2$ plane (see Fig.~2 for the two cases of $N_1=17,N_2=13$ and $N_1=27,N_2=3$). With the further assumption that each individual in the same group pays equally, $x_{ij(\mathrm{YZ})} = X_{i(\mathrm{YZ})}/N_i$, the individual payoff in group 1 is obtained as
\begin{align}
	\left\{ \begin{array}{l}
		u_{1j(\mathrm{CC})} = \frac{1}{N_1} (\frac{N_1^{\alpha}}{N_1^{\alpha}+N_2^{\alpha}})^2 M , \\
		u_{1j(\mathrm{CD})} = \frac{1}{N_1} (\frac{N_1^{\alpha}}{N_1^{\alpha}+N_2^{\alpha-1}})^2 M , \\
		u_{1j(\mathrm{DC})} = \frac{1}{N_1} \frac{N_1^{\alpha-1}}{N_1^{\alpha-1}+N_2^{\alpha}} (1- \frac{1}{N_1} \frac{N_2^{\alpha}}{N_1^{\alpha-1}+N_2^{\alpha}}) M , \\
		u_{1j(\mathrm{DD})} = \frac{1}{N_1} \frac{N_1^{\alpha-1}}{N_1^{\alpha-1}+N_2^{\alpha-1}} (1- \frac{1}{N_1} \frac{N_2^{\alpha-1}}{N_1^{\alpha-1}+N_2^{\alpha-1}}) M , \\
	\end{array} \right.
\end{align}
whereas the expression for group 2 is obtained by replacing 1 with 2 (see Supporting Information).

\section*{Payoff distribution}
From Eq.~4, we can compare the individual payoff in each group optimized in the four cases $(\mathrm{CC}), (\mathrm{CD}), (\mathrm{DC})$, and $(\mathrm{DD})$. For example, the individual payoff in each case is plotted as a function of $N_1$ in Fig.~3 for $N=30$ and $\alpha=2.5$. From Fig.~2, we can see how the restricted amount of resource is distributed between the two groups for each of the four cases $(\mathrm{CC}), (\mathrm{CD}), (\mathrm{DC})$, and $(\mathrm{DD})$.

First, $u_{1j} \geq u_{2j}$ holds for $(\mathrm{CC}), (\mathrm{CD})$, and $(\mathrm{DD})$. For group 1 (with a larger group size), more costs can be recovered with the cooperation policy than the defection policy, so that the former is more advantageous. Next, if the two groups adopt the same policy, the group size power works for group 1 so it can amplify its costs and obtain a higher payoff than a small group. Thus, for the cases $(\mathrm{CC}), (\mathrm{CD})$, and $(\mathrm{DD})$, individuals in group 1 always receive higher payoffs than those in group 2 for any $N_1>N_2$.

For $(\mathrm{DC})$, however, the merit of group size and the demerit in the policy counterbalance each other, so that whether $u_{1j} \geq u_{2j}$ holds depends on $N_1$. With a large asymmetry between group sizes ($N_1 \gg N_2$), the group size merit outweighs the policy demerit so that group 1 members receive higher payoffs. However, group 2 members receive higher payoffs when group size asymmetry is small, ($N_1 \simeq N_2$).

Second, we discuss the dependence of individual payoff $u_{1j}$ in the larger group on $N_1$. As mentioned for $(\mathrm{CC}), (\mathrm{CD}), (\mathrm{DD})$, and $(\mathrm{DC})$, with $N_1 \gg N_2$, group 1 gets a larger payoff. In these cases, there exists an optimal $N_1$ that maximizes $u_{1j}$. This is understood as follows: With the increase in group size, the group payoff increases with $N_1^{\alpha}$ as long as $N_1$ is not too large. The individual payoff, i.e., the group payoff divided by $N_1$, then increases with the group size, as long as $\alpha>1$. As the group size is further increased, however, the increase in the group payoff $U_{1}$ starts to be saturated (see Eq.~3), so that the individual payoff starts to decrease (see Fig.~3-A). Thus, there is an optimal group size for the dominant group (group 1), denoted by $N_1^{\mathrm{op}}$, in spite of the reward increase with $N_1^{\alpha}$. However, in the case of $(\mathrm{DC})$ with $N_1 \simeq N_2$, the reward of group 1 is smaller than that of group 2. Then, the increase in payoff with the group size is not saturated, so that $u_{1j}$ monotonically increases with its group size $N_1$.

The dependence of individual payoff $u_{2j}$ on $N_2$ is studied similarly. For $(\mathrm{CC}), (\mathrm{CD}), (\mathrm{DD})$, and $(\mathrm{DC})$, with $N_1 \gg N_2$, group 2 is dominated by group 1. Therefore, the individual payoff $u_{2j}$ monotonically increases with $N_2$ (i.e., it decreases with $N_1$). For $(\mathrm{DC})$, however, group 2 dominates in the resource competition, and $u_{2j}$ is maximized at an optimal group size (see Fig.~3-B).

Last, we study the total payoff $u_{\mathrm{ave}}$ averaged over the whole population. From Eq.~2, the average payoff decreases when the average cost in all groups increases, which increases as the competition between the groups is stronger. Now, for $(\mathrm{CC}), (\mathrm{CD})$, and $(\mathrm{DD})$, $u_{\mathrm{ave}}$ increases and the average decreases monotonically with $N_1$. In contrast, $u_{\mathrm{ave}}$ has a local minimum for $(\mathrm{DC})$. Therefore, group 1 has a greater advantage than group 2 for $N_1 \gg N_2$, where the competition is weak. However, the payoff for group 2 increases with a decrease in $N_1$, where the competition is stronger. Therefore, $u_{\mathrm{ave}}$ has a local minimum at the intermediate value of $N_1$.

\section*{Hierarchical prisoner's dilemma}

$X_{1(\mathrm{C})}(X_2) > X_{1(\mathrm{D})}(X_2)$ and $X_{2(\mathrm{C})}(X_1) > X_{2(\mathrm{D})}(X_1)$ follow from the above result, so that both the groups should choose the cooperation policy if the other group's cost is constant. Therefore, both groups attempt to choose $(\mathrm{CC})$ policies on their own. However, when both groups' policies are $(\mathrm{CC})$, the resultant payoff is sometimes smaller than that by the choice of $(\mathrm{DD})$ policies. In other words, there exists a prisoner's dilemma in the game of the two groups. (Note that the role of C and D is converted from the standard prisoner's dilemma.) To discuss the existence of the dilemma, we examine the individual payoff matrix for the four cases $(\mathrm{CC})$, $(\mathrm{CD})$, $(\mathrm{DC})$, and $(\mathrm{DD})$ as follows.

\begin{figure}
\begin{center}
\includegraphics[width=0.9\textwidth]{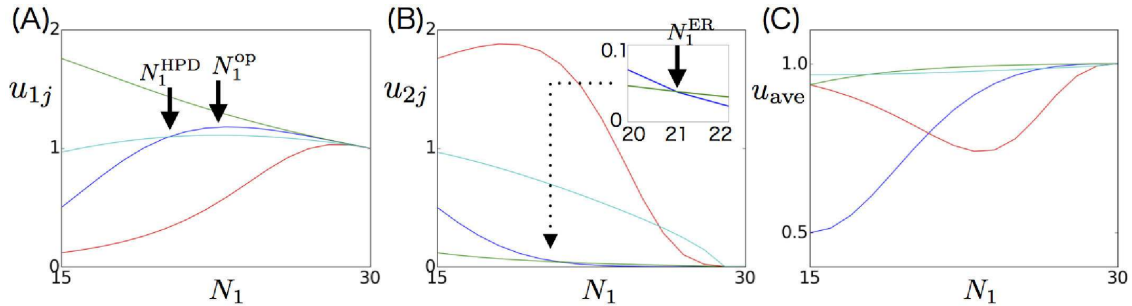}
\caption{Individual payoff of larger group $u_{1j}$ (A) and smaller group $u_{2j}$ (B), and the average payoff over total population $u_{\mathrm{ave}}$ (C) plotted as a function of $N_1$ for four policy scenarios $\{ (\mathrm{CC:blue}), (\mathrm{CD:green}), (\mathrm{DC:red}), (\mathrm{DD:cyan}) \}$ upon $N=30$.
We can see the following four properties from these cases.
For $(\mathrm{CC}), (\mathrm{CD}), and (\mathrm{DD})$, $u_{1j} > u_{2j}$ always holds.
For $(\mathrm{DC})$, $u_{1j} > u_{2j}$ holds in $N_1 \gg N_2$ while $u_{1j} < u_{2j}$ holds in $N_1 \simeq N_2$.
From (A), we note that $u_{1j}$ always has a local maximum.
From (B), we note that $u_{2j}$ has a local maximum for $(\mathrm{DC})$.
From (C), we see that $u_{\mathrm{ave}}$ monotonically increases with $N_1$ for $(\mathrm{CC}), (\mathrm{CD}), and (\mathrm{DD})$, while it has a local minimum for $(\mathrm{DC})$.
}
\end{center}
\end{figure}

First, we consider a case in which the asymmetry between group sizes is small enough, $N_1 \simeq N_2$ (i.e., $N_1 < N_1^{\mathrm{HPD}}$). Here, as seen in Fig.~3, both $u_{1j(\mathrm{CY})} > u_{1j(\mathrm{DY})}$ and $u_{2j(\mathrm{YC})} > u_{2j(\mathrm{YD})}$ hold for any $\mathrm{Y} \in \{ \mathrm{C}, \mathrm{D} \}$. In other words, individuals in each group receive higher payoffs when they cooperate rather than defect regardless of the other group's policy. However, when we compare $(\mathrm{CC})$ and $(\mathrm{DD})$, $X_{i(\mathrm{CC})} > X_{i(\mathrm{DD})}$ holds (see e.g. Fig.~2-A). As group size asymmetry is small, the reward in Eq.~1 does not increase much with a cost. Since the numerator and denominator increase almost at an equal rate, and the payoff is (reward - cost), $u_{ij(\mathrm{DD})} > u_{ij(\mathrm{CC})}$ holds (see Fig.~3). These results indicate a prisoner's dilemma, which we call the hierarchical prisoner's dilemma.

As for the analogy to the military game between the two countries, each country tends to collect more taxes for arms race. However, this increases the loss by war. In contrast, defection (i.e., refusal to pay taxes) will decrease the military cost, and will result in benefit to both countries.

This dilemma does not exist when the asymmetry between group sizes is too large, $N_1 \gg N_2$ (i.e., $N_1 > N_1^{\mathrm{HPD}}$). Indeed, from Fig.~3, $u_{1j(\mathrm{DD})} < u_{1j(\mathrm{CC})}$ conversely holds. This is explained as follows: As seen in Fig.~2-B, $X_{1(\mathrm{CC})} > X_{1(\mathrm{DD})}$ holds as in the case of $N_1 \simeq N_2$. Then, for $N_1 \gg N_2$, the increase in $X_1$ leads to a decrease in $X_2$, resulting in $X_{2(\mathrm{CC})} < X_{2(\mathrm{DD})}$ in contrast to the case of $N_1 \simeq N_2$. In the case of $(\mathrm{CC})$, therefore, group 1 receives a higher payoff than $(\mathrm{DD})$. Thus, when $N_1 \gg N_2$ holds, the hierarchical prisoner's dilemma does not exist. As explained above, whether the hierarchical prisoner's dilemma exists or not can be judged by the configuration between $X_{i(\mathrm{CC})}$ and $X_{i(\mathrm{DD})}$ (compare Fig.~2-A with 2-B).

\section*{Excessive response}
When the hierarchical prisoner's dilemma is avoided for $N_1 \gg N_2$, group 2 receives a higher payoff by following the defection policy. Indeed, whether $u_{2j(\mathrm{YC})}$ or $u_{2j(\mathrm{YD})}$ is larger depends on group size differences, as shown in Fig.~3, which is summarized in Table 1: It shows that $u_{2j(\mathrm{CC})} > u_{2j(\mathrm{CD})}$ holds for $N_1 \simeq N_2$ (i.e., $N_1 < N_1^{\mathrm{ER}}$), while $u_{2j(\mathrm{CC})} < u_{2j(\mathrm{CD})}$ holds for $N_1 \gg N_2$ (i.e., $N_1 > N_1^{\mathrm{ER}}$). In other words, group 2 should choose the defection policy when its group size is small.

\begin{table}[H]
\caption{Dependence of group payoff configurations upon $N_1$ ($N$ is set to 30).}
\begin{center}
\begin{tabular}{ccc}
	\hline
	dependence on $N_1$ & group 1:C & group 1:D \\
	\hline
	\hline
	$15 \le N_1 \le 20$ & $u_{2j(\mathrm{CC})} > u_{2j(\mathrm{CD})}$ & $u_{2j(\mathrm{DC})} > u_{2j(\mathrm{DD})}$ \\
	\hline
	$21 \le N_1 \le 25$ & $u_{2j(\mathrm{CC})} < u_{2j(\mathrm{CD})}$ & $u_{2j(\mathrm{DC})} > u_{2j(\mathrm{DD})}$ \\
	\hline
	$26 \le N_1 \le 29$ & $u_{2j(\mathrm{CC})} < u_{2j(\mathrm{CD})}$ & $u_{2j(\mathrm{DC})} < u_{2j(\mathrm{DD})}$ \\
	\hline
\end{tabular}
\end{center}
\end{table}

We now explain the reason for this unexpected outcome. In Fig.~2-B, $X_{2(\mathrm{YC})} \simeq X_{2(\mathrm{YD})}$ and $X_{1(\mathrm{YC})} > X_{1(\mathrm{YD})}$ hold. When group 2 changes its policy from D to C, its cost could increase if $X_1$ is fixed (see the broken blue line in Fig.~2-B). However, in response to the policy of group 2, group 1 will increase its cost, so that group 1 takes back the resource that could be lost with the cost increase in group 2. This response from group 1 is excessive for group 2, and the cost increase in group 2 is suppressed so that $X_{2(\mathrm{YC})} \simeq X_{2(\mathrm{YD})}$ holds (see the cross-point $(\mathrm{CC})$ in Fig.~2-B). Recall that the reward of group 2 is given by $\frac{X_2N_2^{\alpha}}{X_1N_1^{\alpha}+X_2N_2^{\alpha}} M$ according to Eq.~1. Then, with the increase in $X_1$, the reward of group 2 decreases as long as its cost does not increase much. Therefore, $u_{2j(\mathrm{CY})} < u_{2j(\mathrm{DY})}$ holds. As explained above, whether this ``excessive response" occurs is determined by the positional relationship between $X_{i(\mathrm{YC})}$ and $X_{i(\mathrm{YD})}$, which depends on $N_1/N$.

In summary, an excessive response by group 1 against the cooperation policy of group 2 leads to this unexpected outcome. When individuals in group 2 play only the intra-group game, they attempt to cooperate. By taking the inter-group game into account, they can avoid an excessive response from the other group against the cooperative strategy, so that they choose the defection policy.

Let us recall the analogy of the present game to the military game between two countries, where the individual cost corresponds to the military tax of the people, and cooperation means the armament. If a small country collects more tax for arms race, the larger country increases its armaments in response, so that the payoff of the smaller country decreases. Hence, defection, i.e., decreasing the military tax, is a better policy for the smaller country.

\section*{Dependence on $N$}
The excessive response and hierarchical prisoner's dilemma discussed so far occur for any total population $N$, while the region in which the former exists (defined as $N_1 > N_1^{\mathrm{ER}}$) decreases with an increase in $N$.
As shown in Supporting Information, the value $R_{\mathrm{ER}}:= N_1^{\mathrm{ER}}/N$ is estimated as
\begin{align}
	\nonumber
	&  \ R_{\mathrm{ER}} \simeq 1 - N^{-\frac{1}{\alpha+1}},
\end{align}
in the limit of $N \rightarrow \infty$.
However, the hierarchical prisoner's dilemma occurs in the region $N_1 < N_1^{\mathrm{HPD}}$, and the value $R_{\mathrm{HPD}} := N_1^{\mathrm{HPD}}/N$ satisfies, in the limit of $N \rightarrow \infty$,
\begin{align}
	\nonumber
	& (1 + K^{\alpha})^2 = 1 + K^{\alpha-1},
\end{align}
with $K := (1-R_{\mathrm{HPD}})/R_{\mathrm{HPD}}$.

As for the optimal size $N_1^{\mathrm{op}}$ for the policies $\mathrm{CC}$, $R_{\mathrm{op}} := N_1^{\mathrm{op}}/N$ satisfies
\begin{align}
	\nonumber
	& 2\alpha(1-R_{\mathrm{op}})^{\alpha-1} = R_{\mathrm{op}}^{\alpha}+(1-R_{\mathrm{op}})^{\alpha}
\end{align}
in the limit of $N \rightarrow \infty$ (see Supporting Information).

\section*{Dependence on $\alpha$}

\begin{figure}
\begin{center}
\includegraphics[width=0.4\textwidth]{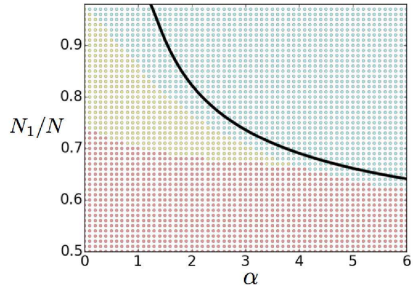}
\caption{The regimes with the hierarchical prisoner's dilemma (red) and excessive response (cyan) plotted against $\alpha$ (horizontal axis) and $N_1/N$ (vertical axis). In the yellow region, neither of the two exists. When $\alpha > 2$ holds, there exists optimal group size $N_1^{\mathrm{op}}$ (black line) for the larger group. Computed for $N=50$.
}
\end{center}
\end{figure}

So far, we have adopted $\alpha=2.5$. 
Indeed, the behaviors we reported here are universally observed as long as $\alpha > 0$.
The phase diagram for the regions with the hierarchical prisoner's dilemma and excessive response are shown in Fig.~4 against the change in $\alpha$ and $N_1/(N_1+N_2)$. Since $\alpha$ represents the advantage of the larger group relative to the smaller group, the region with excessive response (HPD) decreases (increases) with a decrease in $\alpha$.

For $\alpha <2$, $u_{2j(\mathrm{DD})} > u_{1j(\mathrm{DD})}$ always holds, and for $\alpha < 1.5$, $u_{2j(\mathrm{DC})} > u_{1j(\mathrm{DC})}$ holds. In other words, individuals in the smaller group receive higher payoffs than those in the larger group, when the latter's policy is defection. For details of the behavior for $\alpha < 2$, see Supporting Information.

\section*{Discussion}
In the present paper, we have introduced and analyzed a new game, called the HPG game. A cost is paid according to each player's strategy, while groups play a game to compete for resources, depending on the collected costs in a group. The policies of the group are determined as cooperation (C) or defection (D) according to whether each player in the group pays a cost or not.

In this game, we uncovered a novel type of dilemma, which we term the hierarchical prisoner's dilemma: Each group receives a higher payoff by choosing policy C rather than D if the other group's cost is constant. Hence $(\mathrm{CC})$ is the Nash equilibrium in the game between the groups. However, the payoff that each group receives is lower than if the groups chose $(\mathrm{DD})$, and thus the prisoner's dilemma occurs. We again emphasize that C (D) means cooperation (defection) in terms of the intra-group game, while C (D) inversely implies defection (cooperation) in terms of the inter-group game.

However, for each individual, there also exists a dilemma, typical of the PG game. Every individual prefers not to pay the cost, i.e., to choose D. It is interesting and important that in our HPG game, the cooperation in intra-group game leads to defection in the inter-group game. This hierarchical prisoner's dilemma was first formulated in the present study.

In the present model, this dilemma exists if the two group sizes are not very different. As the group size difference increases, the larger group always gets an advantage by paying the costs of competition and, thus, avoiding this hierarchical prisoner's dilemma. If the difference is sufficiently large, both groups receive higher payoffs when the smaller group follows the D policy. Indeed, if the smaller group pursues C, i.e., it pays more to compete with the larger group, the latter pays more, too, so that competition for resources increases, and the smaller group suffers a loss, as a result of what is here termed excessive response. Hence, the smaller group abandons the cost competition, so that the dilemma is avoided.

These findings may have some implications on the arms race or struggle between groups. If the two groups or countries are not much different in size, they cannot avoid an arms race, which is costly for both members. With cooperative members, the race would be stronger, and the costs larger. In contrast, when the two countries or groups are quite different in size, the smaller group would abandon the race, averting the loss. Of course, in reality, the interaction between groups and resource allocation to each is more complicated, and choice of policy in each group is not simply determined by the actions of individuals. Nevertheless, the group size dependence of the dilemma and the choice of policies may be relevant to understand the real society.

In the present paper, we assumed, for simplicity, that all individuals in a group follow the same action. In reality, the cost each individual within the same group pays can be different. Indeed, as individuals play the PG game in each group, there exists a prisoner's dilemma within the group. In fact, defectors free-ride on cooperators, so the group would be defeated with their increase, as shown in multi-level selection studies\cite{Traulsen2006}. In the present HPG, however, if the fraction of free-riders increases, the cost of the struggle between the groups decreases. Thus, the payoff could increase, depending also on each group size. Hence, it will be interesting to discuss the distribution of costs and payoffs within a group, together with its dependence on group size.

The significant role of group size uncovered here is related to Wrangham's power-of-imbalance hypothesis [18], in which animals attempt to form larger groups to dominate other smaller groups when the available resources are limited. Given that a larger group has an advantage, we have found an optimal group size that maximizes the payoff of the larger group. This is in contrast to the naive expectation that larger groups that would ultimately coalesce to a single group would be more advantageous. Counterintuitively, an optimal group size exists because of the limitation of available resources, which causes the hierarchical prisoner's dilemma, and the size depends on the degree of power imbalance between groups. An investigation into the appropriate size of animal groups in nature might provide some insight. Extension of the present game to many groups with introduction of migration among groups as well as population dynamics as in dynamical system game\cite{Akiyama2000} should be important for future issue of an appropriate size distribution.

\begin{acknowledgments}
The authors would like to thank S. Sasa, N. Saito, and T. S. Hatakeyama for useful discussions. This research is partially supported by the Platform for Dynamic Approaches to Living System of the Japan Agency for Medical Research and Development(AMED).
\end{acknowledgments}


\newpage

\setcounter{figure}{0}
\renewcommand{\figurename}{FIG. S}

\begin{center}
{\LARGE {\bf Supporting Information}}
\end{center}

\section{Game between individuals}

In this section, we derive various functions in ``Game between individuals" in our paper.

In our model, a payoff function of individual $j \in \{ 1, \cdots , N_i \}$ in group $i \in \{ 1,2 \}$ is given by
\begin{align}
	& u_{ij} = \frac{1}{N_i} \frac{X_iN_i^{\alpha}}{\sum_i X_iN_i^{\alpha}} M - x_{ij} .
\end{align}
Here, $X_i (= \sum_j x_{ij})$ is the total cost in group $i$.
From Eq.~1, the total payoff in group $i$ is given by
\begin{align}
	& U_i = \frac{X_iN_i^{\alpha}}{\sum_i X_iN_i^{\alpha}} M - X_i .
\end{align}

We now consider how much each individual in group 1 pays. Here, the total cost in the opponent group $X_2$ is fixed. First, we consider a case in which individuals within group 1 defect. Then, each of them chooses his or her own cost $x_{1j(\mathrm{D})}(X_2)$ to maximize one's own payoff $u_{1j}$.
\begin{align}
	\nonumber
	& \left. \frac{\partial u_{1j}}{\partial x_{1j}} \right|_{\{ x_{1j}=x_{1j(\mathrm{D})}(X_2) | j \in \{ 1, \cdots , N_1 \} \}} = 0 \quad (\forall j)\\
	\nonumber
	& \Leftrightarrow \frac{1}{N_1} \frac{X_2N_2^{\alpha}N_1^{\alpha}}{(X_{1(\mathrm{D})}(X_2)N_1^{\alpha}+X_2N_2^{\alpha})^2} M - 1 = 0 \\
	\nonumber
	& \Leftrightarrow X_{1(\mathrm{D})}(X_2) = \sqrt{\frac{M}{N_1} \left(\frac{N_2}{N_1}\right)^{\alpha} X_2} - \left(\frac{N_2}{N_1}\right)^{\alpha}X_2
\end{align}
Here, we define $X_{1(\mathrm{D})}(X_2) = \sum_j x_{1j(\mathrm{D})}(X_2)$. 
In this way, the condition for optimization of each individual cost is determined only by the total cost in the group. 
From Eq.~2, the total payoff in group 1 $U_{1(\mathrm{D})}(X_2)$ is determined by
\begin{align}
	\nonumber
	U_{1(\mathrm{D})}(X_2) & = \frac{X_{1(\mathrm{D})}(X_2)N_1^{\alpha}}{X_{1(\mathrm{D})}(X_2)N_1^{\alpha}+X_2N_2^{\alpha}} M - X_{1(\mathrm{D})}(X_2) \\
	& = M - \left(\sqrt{N_1} + \frac{1}{\sqrt{N_1}}\right) \sqrt{M\left(\frac{N_2}{N_1}\right)^{\alpha}X_2} + \left(\frac{N_2}{N_1}\right)^{\alpha}X_2 .
\end{align}

Second, we consider a case in which individuals cooperate within group 1. Then, they collect a cost of $X_{1(\mathrm{C})}(X_2)$ to maximize the group's total payoff $U_1$.
\begin{align}
	\nonumber
	& \frac{\partial U_1}{\partial X_1}|_{X_1 = X_{1(\mathrm{C})}(X_2)} = 0 \\
	\nonumber
	& \Leftrightarrow \frac{X_2N_2^{\alpha}X_1^{\alpha}}{X_{1(\mathrm{C})}(X_2)N_1^{\alpha}+X_2N_2^{\alpha}} M - 1 = 0 \\
	\nonumber
	& \Leftrightarrow X_{1(\mathrm{C})}(X_2) = \sqrt{M \left(\frac{N_2}{N_1}\right)^{\alpha} X_2} - \left(\frac{N_2}{N_1}\right)^{\alpha}X_2
\end{align}
From Eq.~2, the total payoff in group 1 $U_{1(\mathrm{C})}(X_2)$ is determined by
\begin{align}
	\nonumber
	U_{1(\mathrm{C})}(X_2) & = \frac{X_{1(\mathrm{C})}(X_2)N_1^{\alpha}}{X_{1(\mathrm{C})}(X_2)N_1^{\alpha}+X_2N_2^{\alpha}} M - X_{1(\mathrm{C})}(X_2) \\
	& = M - 2 \sqrt{M\left(\frac{N_2}{N_1}\right)^{\alpha}X_2} + \left(\frac{N_2}{N_1}\right)^{\alpha}X_2 .
\end{align}
We now check whether $X_{1(\mathrm{C})}(X_2)$ or $X_{1(\mathrm{D})}(X_2)$ is larger. From $N_1 \ge 1$, $X_{1(\mathrm{C})}(X_2) \ge X_{1(\mathrm{D})}(X_2)$ holds for any $X_2$. In addition, we check the magnitude relation between $U_{1(\mathrm{C})}(X_2)$ and $U_{1(\mathrm{D})}(X_2)$. The first and third terms are equal between Eq.~3 and Eq.~4. Then, by comparing the second terms with the use of
\begin{align}
	\nonumber
	& \sqrt{N_1} + \frac{1}{\sqrt{N_1}} \ge 2
\end{align}
we get $U_{1(\mathrm{C})}(X_2) \ge U_{1(\mathrm{D})}(X_2)$ for any $X_2$.

The above results for group 1 also hold for group 2, and we obtain
\begin{align}
	\nonumber
	& X_{2(\mathrm{D})}(X_1) = \sqrt{\frac{M}{N_2} \left(\frac{N_1}{N_2}\right)^{\alpha} X_1} - \left(\frac{N_1}{N_2}\right)^{\alpha}X_1 \\
	\nonumber
	& U_{2(\mathrm{D})}(X_1) = M - \left(\sqrt{N_2} + \frac{1}{\sqrt{N_2}}\right) \sqrt{M\left(\frac{N_1}{N_2}\right)^{\alpha}X_1} + \left(\frac{N_1}{N_2}\right)^{\alpha}X_1 \\
	\nonumber
	& X_{2(\mathrm{C})}(X_1) = \sqrt{M \left(\frac{N_1}{N_2}\right)^{\alpha} X_1} - \left(\frac{N_1}{N_2}\right)^{\alpha}X_1 \\
	\nonumber
	& U_{2(\mathrm{C})}(X_1) = M - 2 \sqrt{M\left(\frac{N_1}{N_2}\right)^{\alpha}X_1} + \left(\frac{N_1}{N_2}\right)^{\alpha}X_1.
\end{align}
Then, both $X_{2(\mathrm{C})}(X_1) \ge X_{2(\mathrm{D})}(X_1)$ and $U_{2(\mathrm{C})}(X_1) \ge U_{2(\mathrm{D})}(X_1)$ also hold.

We now examine whether individuals in each group are cooperators or defectors. Then, four kinds of equilibrium points are defined as $(\mathrm{CC}), (\mathrm{CD}), (\mathrm{DC}), (\mathrm{DD})$. Here, the left index indicates the group 1 policy, while the right index indicates the group 2 policy. Then, the group's total cost in each of four equilibrium points is given by
\begin{align}
	\nonumber
	& \left\{ \begin{array}{l}
		X_{1(\mathrm{YZ})} = X_{1(\mathrm{Y})}(X_{2(\mathrm{YZ})}) \\
		X_{2(\mathrm{YZ})} = X_{1(\mathrm{Z})}(X_{1(\mathrm{YZ})}) \\
	\end{array} \right.
\end{align}
Here, $\mathrm{Y} \in \{ \mathrm{C,D} \}$ and $\mathrm{Z} \in \{ \mathrm{C,D} \}$ hold.
For example, when $\mathrm{Y}=\mathrm{C}$ and $\mathrm{Z}=\mathrm{D}$ hold, $X_{1(\mathrm{CD})}$ and $X_{2(\mathrm{CD})}$ are given by
\begin{align}
	\nonumber
	& \left\{ \begin{array}{l}
		X_{1(\mathrm{CC})} = X_{1(\mathrm{C})}(X_{2(\mathrm{CC})}) \\
		X_{2(\mathrm{CC})} = X_{2(\mathrm{C})}(X_{1(\mathrm{CC})}) \\
	\end{array} \right. \\
	\nonumber
	& \Leftrightarrow \left\{ \begin{array}{l}
		X_{1(\mathrm{CC})} = \sqrt{M (\frac{N_2}{N_1})^{\alpha} X_{2(\mathrm{CC})}} - (\frac{N_2}{N_1})^{\alpha}X_{2(\mathrm{CC})} \\
		X_{2(\mathrm{CC})} = \sqrt{M (\frac{N_2}{N_1})^{\alpha} X_{2(\mathrm{CC})}} - (\frac{N_2}{N_1})^{\alpha}X_{2(\mathrm{CC})} \\
	\end{array} \right. \\
	\nonumber
	& \Leftrightarrow X_{1(\mathrm{CC})} = X_{2(\mathrm{CC})} = \frac{N_1^{\alpha}N_2^{\alpha}}{(N_1^{\alpha}+N_2^{\alpha})^2} M .
\end{align}
Also, we obtain
\begin{align}
	\nonumber
	& \left\{ \begin{array}{l}
		X_{1(\mathrm{CD})} = \frac{N_1^{\alpha}N_2^{\alpha-1}}{(N_1^{\alpha}+N_2^{\alpha-1})^2} M \\
		X_{2(\mathrm{CD})} = \frac{1}{N_2} \frac{N_1^{\alpha}N_2^{\alpha-1}}{(N_1^{\alpha}+N_2^{\alpha-1})^2} M \\
	\end{array} \right. \\
	\nonumber
	& \left\{ \begin{array}{l}
		X_{1(\mathrm{DC})} = \frac{1}{N_1} \frac{N_1^{\alpha-1}N_2^{\alpha}}{(N_1^{\alpha-1}+N_2^{\alpha})^2} M \\
		X_{2(\mathrm{DC})} = \frac{N_1^{\alpha-1}N_2^{\alpha}}{(N_1^{\alpha-1}+N_2^{\alpha})^2} M \\
	\end{array} \right. \\
	\nonumber
	& \left\{ \begin{array}{l}
		X_{1(\mathrm{DD})} = \frac{1}{N_1} \frac{N_1^{\alpha-1}N_2^{\alpha-1}}{(N_1^{\alpha-1}+N_2^{\alpha-1})^2} M \\
		X_{2(\mathrm{DD})} = \frac{1}{N_2} \frac{N_1^{\alpha-1}N_2^{\alpha-1}}{(N_1^{\alpha-1}+N_2^{\alpha-1})^2} M . \\
	\end{array} \right.
\end{align}
From the assumption that all individuals pay equally within a group, the individual cost, denoted as $x_{ij(\mathrm{YZ})}$, is given by
\begin{align}
	\nonumber
	& x_{ij(\mathrm{YZ})} = X_{i(\mathrm{YZ})}/N_i
\end{align}
Then, from $X_{1(\mathrm{YZ})}$ and $X_{2(\mathrm{YZ})}$, each of the groups' payoffs, denoted as $U_{1(\mathrm{YZ})}$ and $U_{2(\mathrm{YZ})}$, is given by
\begin{align}
	& \left. \begin{array}{l}
		\left\{ \begin{array}{l}
			U_{1(\mathrm{CC})} = (\frac{N_1^{\alpha}}{N_1^{\alpha}+N_2^{\alpha}})^2 M \\
			U_{2(\mathrm{CC})} = (\frac{N_2^{\alpha}}{N_1^{\alpha}+N_2^{\alpha}})^2 M \\
		\end{array} \right. \\
		\left\{ \begin{array}{l}
			U_{1(\mathrm{CD})} = (\frac{N_1^{\alpha}}{N_1^{\alpha}+N_2^{\alpha-1}})^2 M \\
			U_{2(\mathrm{CD})} = \frac{N_1^{\alpha-1}}{N_1^{\alpha}+N_2^{\alpha-1}} (1-\frac{1}{N_2}\frac{N_1^{\alpha}}{N_1^{\alpha}+N_2^{\alpha-1}}) M \\
		\end{array} \right. \\
		\left\{ \begin{array}{l}
			U_{1(\mathrm{DC})} = \frac{N_1^{\alpha-1}}{N_1^{\alpha-1}+N_2^{\alpha}} (1-\frac{1}{N_1}\frac{N_2^{\alpha}}{N_1^{\alpha-1}+N_2^{\alpha}}) M \\
			U_{2(\mathrm{DC})} = (\frac{N_2^{\alpha}}{N_1^{\alpha-1}+N_2^{\alpha}})^2 M \\
		\end{array} \right. \\
		\left\{ \begin{array}{l}
			U_{1(\mathrm{DD})} = \frac{N_1^{\alpha-1}}{N_1^{\alpha-1}+N_2^{\alpha-1}} (1-\frac{1}{N_1}\frac{N_2^{\alpha-1}}{N_1^{\alpha-1}+N_2^{\alpha-1}}) M \\
			U_{2(\mathrm{DD})} = \frac{N_2^{\alpha-1}}{N_1^{\alpha-1}+N_2^{\alpha-1}} (1-\frac{1}{N_2}\frac{N_1^{\alpha-1}}{N_1^{\alpha-1}+N_2^{\alpha-1}}) M . \\
		\end{array} \right.
	\end{array} \right.
\end{align}
Individual payoff, denoted as $u_{ij(\mathrm{YZ})}$, is straightforwardly obtained by
\begin{align}
	\nonumber
	& u_{ij(\mathrm{YZ})} = U_{i(\mathrm{YZ})}/N_i.
\end{align}

\section{Dependence on $N$}

In this section, we consider under what condition each of two phenomena, excessive response and the hierarchical prisoner's dilemma, appears, given the total population $N$.

First, we consider the region where excessive response happens. Since $u_{1j(\mathrm{CY})} > u_{1j(\mathrm{DY})}$ holds for any $\mathrm{Y} \in \{ \mathrm{C}, \mathrm{D} \}$, group 1 always chooses policy C.
Thus, in order to determine whether group 2 should choose policy C or D, we compare $u_{2j(\mathrm{CC})}$ and $u_{2j(\mathrm{CD})}$, given by
\begin{align}
	\nonumber
	& u_{2j(\mathrm{CC})} = \frac{1}{N(1-R)}\left(\frac{(1-R)^{\alpha}}{R^{\alpha}+(1-R)^{\alpha}}\right)^2 M \\
	\nonumber
	& u_{2j(\mathrm{CD})} = \frac{1}{N(1-R)}\frac{(1-R)^{\alpha-1}}{NR^{\alpha}+(1-R)^{\alpha-1}} \left(1-\frac{1}{N(1-R)}\frac{NR^{\alpha}}{NR^{\alpha}+(1-R)^{\alpha-1}}\right) M
\end{align}
respectively, where we define $R=N_1/N$.
Noting that the condition $R_{\mathrm{ER}}$ at which excessive response appears is given by $u_{2j(\mathrm{CC})} = u_{2j(\mathrm{CD})}$, we get
\begin{align}
	& \left(\frac{(1-R_{\mathrm{ER}})^{\alpha}}{NR_{\mathrm{ER}}^{\alpha}+(1-R_{\mathrm{ER}})^{\alpha}}\right)^2 = \frac{(1-R_{\mathrm{ER}})^{\alpha-1}}{NR_{\mathrm{ER}}^{\alpha}+(1-R_{\mathrm{ER}})^{\alpha-1}} \left(1-\frac{1}{N(1-R_{\mathrm{ER}})}\frac{NR_{\mathrm{ER}}^{\alpha}}{NR_{\mathrm{ER}}^{\alpha}+(1-R_{\mathrm{ER}})^{\alpha-1}}\right) .
\end{align}
Noting that for Eq.~6 to be satisfied against $N \rightarrow \infty$, $(1-R)$ has to diverge with it and, taking the leading order with $N$, we get
\begin{align}
	\nonumber
	& \left( \frac{(1-R_{\mathrm{ER}})^{\alpha}}{NR_{\mathrm{ER}}^{\alpha}} \right)^2 = \frac{(1-R_{\mathrm{ER}})^{\alpha-1}}{NR_{\mathrm{ER}}^{\alpha}}.
\end{align}
Thus, we obtain
\begin{align}
	\nonumber
	& R_{\mathrm{ER}} = 1 - N^{-\frac{1}{\alpha+1}} .
\end{align}

Second, we consider the region where the hierarchical prisoner's dilemma happens.
Since $u_{1j(\mathrm{CY})} > u_{1j(\mathrm{DY})}$ holds for any $\mathrm{Y} \in \{ \mathrm{C}, \mathrm{D} \}$, we compare $u_{1j(\mathrm{CC})}$ and $u_{1j(\mathrm{DD})}$.
$u_{1j(\mathrm{CC})}$ and $u_{1j(\mathrm{DD})}$ are given by
\begin{align}
	\nonumber
	& u_{1j(\mathrm{CC})} = \frac{1}{NR}\left(\frac{R^{\alpha}}{R^{\alpha}+(1-R)^{\alpha}}\right)^2 M , \\
	\nonumber
	& u_{1j(\mathrm{DD})} = \frac{1}{NR}\frac{R^{\alpha-1}}{R^{\alpha-1}+(1-R)^{\alpha-1}} \left(1-\frac{1}{NR}\frac{(1-R)^{\alpha-1}}{R^{\alpha-1}+(1-R)^{\alpha-1}}\right) M .
\end{align}
Noting that $u_{1j(\mathrm{CC})} = u_{1j(\mathrm{DD})}$ holds for $R=R_{\mathrm{HPD}}$, we get
\begin{align}
	\nonumber
	& \left(\frac{R_{\mathrm{HPD}}^{\alpha}}{R_{\mathrm{HPD}}^{\alpha}+(1-R_{\mathrm{HPD}})^{\alpha}}\right)^2 = \frac{R_{\mathrm{HPD}}^{\alpha-1}}{R_{\mathrm{HPD}}^{\alpha-1}+(1-R_{\mathrm{HPD}})^{\alpha-1}} \left(1-\frac{1}{NR_{\mathrm{HPD}}}\frac{(1-R_{\mathrm{HPD}})^{\alpha-1}}{R_{\mathrm{HPD}}^{\alpha-1}+(1-R_{\mathrm{HPD}})^{\alpha-1}}\right) .
\end{align}
In the limit of $N \rightarrow \infty$, we obtain
\begin{align}
	\nonumber
	& (1+K^{\alpha})^2 = 1+K^{\alpha-1}
\end{align}
with $K := (1-R_{\mathrm{HPD}})/R_{\mathrm{HPD}}$.

Third, we consider the optimal size for the larger group in the case of $\mathrm{CC}$.
$u_{1j(\mathrm{CC})}$ is given by
\begin{align}
	\nonumber
	& u_{1j(\mathrm{CC})} = \frac{1}{NR}\left(\frac{R^{\alpha}}{R^{\alpha}+(1-R)^{\alpha}}\right)^2 M .
\end{align}
Noting that $u_{1j(\mathrm{CC})}$ is maximized for $R=R_{\mathrm{op}}$, we get
\begin{align}
	\nonumber
	& \left. \frac{\partial u_{1j(\mathrm{CC})}}{\partial R} \right|_{R=R_{\mathrm{op}}} = 0 \\
	\nonumber
	& \Leftrightarrow \frac{R_{\mathrm{op}}^{2\alpha}(2\alpha(1-R_{\mathrm{op}})^{\alpha-1}-(R_{\mathrm{op}}^{\alpha}+(1-R_{\mathrm{op}})^{\alpha}))}{R_{\mathrm{op}}^2(R_{\mathrm{op}}^{\alpha}+(1-R_{\mathrm{op}})^{\alpha})^3}=0 \\
	\nonumber
	& \Leftrightarrow 2\alpha(1-R_{\mathrm{op}})^{\alpha-1} = R_{\mathrm{op}}^{\alpha}+(1-R_{\mathrm{op}})^{\alpha} .
\end{align}

From the above result, the region of excessive response, in the limit of $N \rightarrow \infty$, decreases according to $N$, while that of hierarchical PD and optimal group size does not change (See Fig.~S1).

\begin{figure}[H]
\begin{center}
\includegraphics[width=0.9\textwidth]{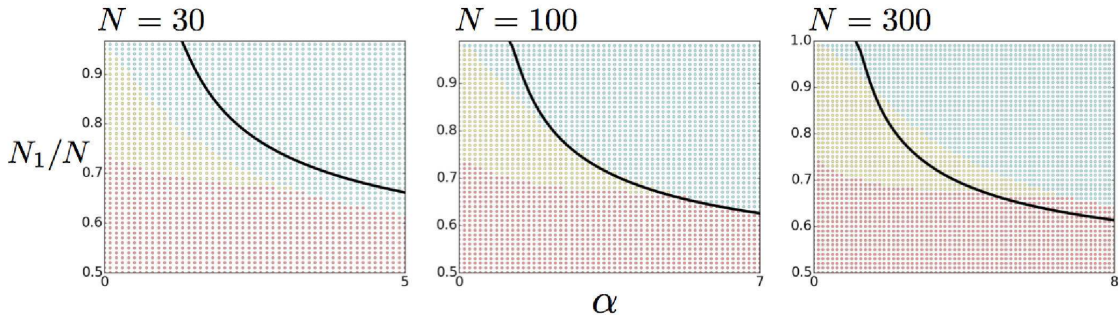}
\caption{The regimes with the hierarchical PD (red) and excessive response (cyan) plotted against $\alpha$ (horizontal axis) and $N_1/N$ (vertical axis). In the yellow region, neither of the two exists. Computed for $N=30$ (left), $N=100$ (central) and $N=300$ (right).
}
\end{center}
\end{figure}

\section{Dependence on $\alpha$}

In our paper, we assume that $\alpha > 2$, which is require in order that individuals in the larger group receive higher payoffs than those in the smaller group when both groups choose the same policy. However, when $\alpha < 2$ holds, individuals in the larger group cannot always receive higher payoffs than those in the smaller group. In this section, we show that several transitions appear with decreasing $\alpha$, where individual payoff in the smaller group is larger than that in the larger group.

In the following, we study the case with $N_2>1$, since the case $N_2=1$ is exceptional as the group is nothing but an individual.
In this case
\begin{align}
	& u_{ij(\mathrm{YC})} = u_{ij(\mathrm{YD})}
\end{align}
holds for any $i \in \{ 1,2 \}$ and $\mathrm{Y} \in \{ \mathrm{C}, \mathrm{D} \}$ as seen in Eq.~5.
Eq.~7 indicates that the individual payoff in the smaller group is independent of its policy, which is identical to the group's payoff. Indeed for $N_2=1$, the group has no ``hierarchy" and, therefore, is not suitable for the present study. As $\alpha$ decreases, so does the advantage of group 1 over 2. With the decrease in $\alpha$, group 2 gets an advantage over 1, first for $\mathrm{DD}$, then for $\mathrm{CD}$, and then for $\mathrm{CC}$, and from $N_1 \gg N_2$ to $N_1 \simeq N_2$ in that order.

First, we recall the case of $\alpha > 2$. In this case, we obtain (see Fig.~S2)
\begin{align}
	\nonumber
	& u_{1j(\mathrm{CC})} > u_{2j(\mathrm{CC})} \\
	\nonumber
	& u_{1j(\mathrm{CD})} > u_{2j(\mathrm{CD})} \\
	\nonumber
	& u_{1j(\mathrm{DC})} \left\{ \begin{array}{ll}
		< u_{2j(\mathrm{DC})} & (N_1 \simeq N_2) \\
		> u_{2j(\mathrm{DC})} & (N_1 \gg N_2) \\
	\end{array} \right. \\
	\nonumber
	& u_{1j(\mathrm{DD})} > u_{2j(\mathrm{DD})} .
\end{align}

\begin{figure}[H]
\begin{center}
\includegraphics[width=0.5\textwidth]{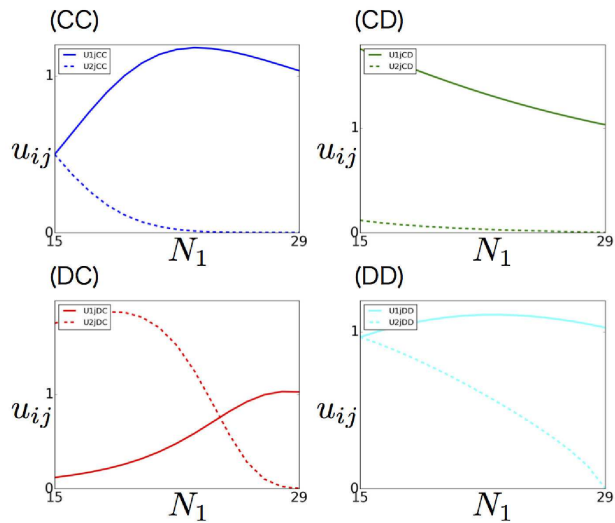}
\caption{Individual payoff $u_{1j(\mathrm{YZ})}$ (solid line) and $u_{2j(\mathrm{YZ})}$ (broken line) plotted as a function of $N_1$ for $N=30, \alpha=2.5$.
The blue (upper left), green (upper right), red (lower left) and cyan (lower right) lines indicate $\mathrm{YZ}=\mathrm{CC}, \mathrm{CD}, \mathrm{DC}$, and $\mathrm{DD}$, respectively.
}
\end{center}
\end{figure}

Second, for $1.5 < \alpha < 2$ (see Fig.~S3),
\begin{align}
	\nonumber
	& u_{1j(\mathrm{CC})} > u_{2j(\mathrm{CC})} \\
	\nonumber
	& u_{1j(\mathrm{CD})} > u_{2j(\mathrm{CD})} \\
	\nonumber
	& u_{1j(\mathrm{DC})} \left\{ \begin{array}{ll}
		< u_{2j(\mathrm{DC})} & (N_1 \simeq N_2) \\
		> u_{2j(\mathrm{DC})} & (N_1 \gg N_2) \\
	\end{array} \right. \\
	\nonumber
	& u_{1j(\mathrm{DD})} < u_{2j(\mathrm{DD})} .
\end{align}

\begin{figure}[H]
\begin{center}
\includegraphics[width=0.5\textwidth]{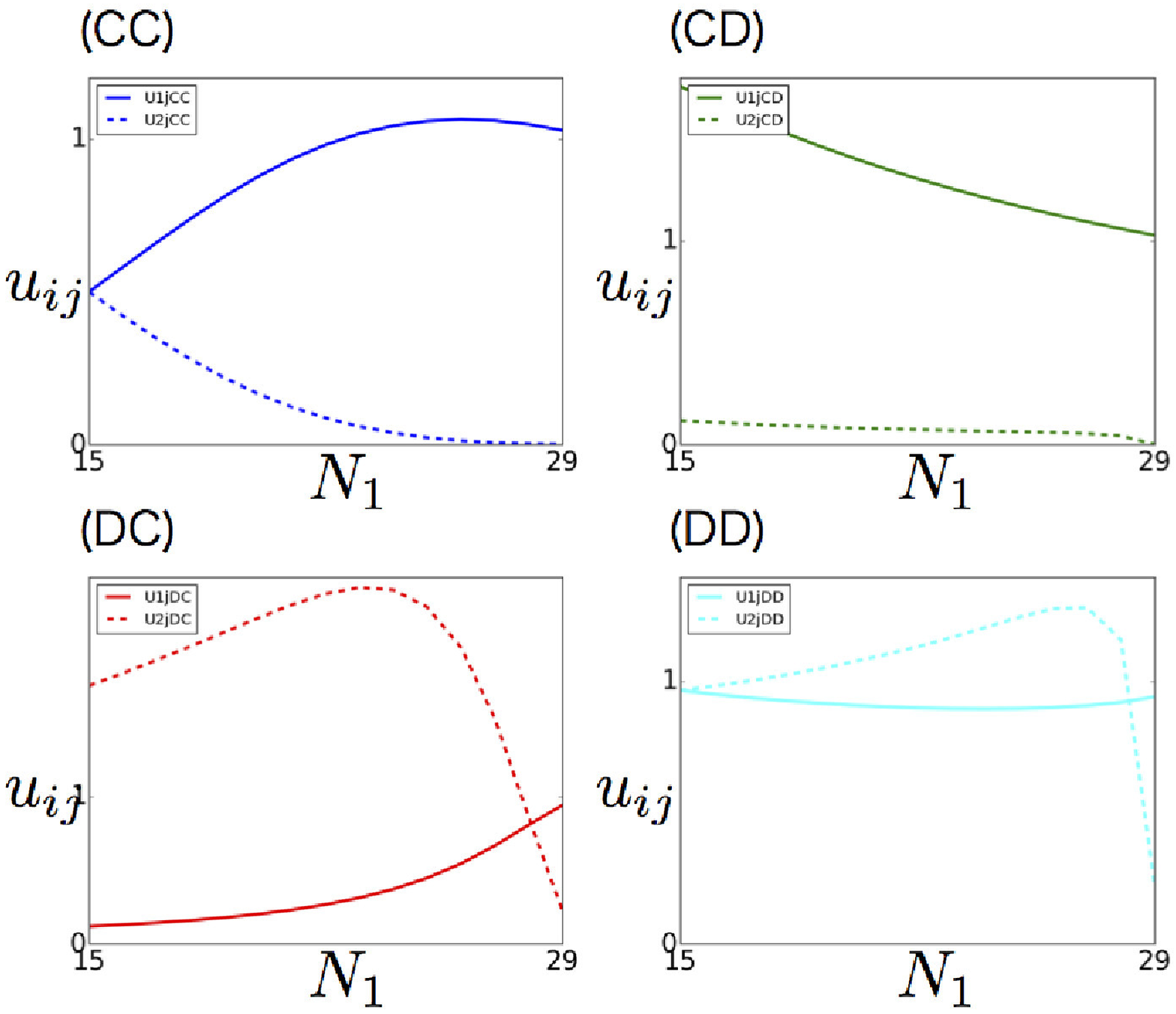}
\caption{Individual payoff $u_{1j(\mathrm{YZ})}$ (solid line) and $u_{2j(\mathrm{YZ})}$ (broken line) plotted as a function of $N_1$ for $N=30, \alpha=1.7$.
The blue (upper left), green (upper right), red (lower left), and cyan (lower right) lines indicates $\mathrm{YZ}=\mathrm{CC}, \mathrm{CD}, \mathrm{DC}$, and $\mathrm{DD}$, respectively.
}
\end{center}
\end{figure}
The difference between $1.5<\alpha<2$ and $\alpha>2$ lies in the behavior of $u_{ij(\mathrm{DD})}$. For $1.5<\alpha<2$, individuals in the smaller group always
\footnote{As an exception, for $N_2=1$, $u_{1j(\mathrm{DD})} > u_{2j(\mathrm{DD})}$ holds. When $\alpha$ is still smaller ($\alpha<1.5$), $u_{1j(\mathrm{DD})} < u_{2j(\mathrm{DD})}$ holds even for $N_2=1$.}
receive higher payoffs than those in the larger group do in the case of $\mathrm{DD}$.
Here, $u_{1j(\mathrm{DD})}$ and $u_{2j(\mathrm{DD})}$ are given by
\begin{align}
	\nonumber
	& u_{1j(\mathrm{DD})} = \frac{1}{N_1}\frac{N_1^{\alpha-1}}{N_1^{\alpha-1}+N_2^{\alpha-1}} \left(1-\frac{1}{N_1}\frac{N_1^{\alpha-1}}{N_1^{\alpha-1}+N_2^{\alpha-1}}\right) M \\
	\nonumber
	& u_{2j(\mathrm{DD})} = \frac{1}{N_2}\frac{N_2^{\alpha-1}}{N_1^{\alpha-1}+N_2^{\alpha-1}} \left(1-\frac{1}{N_2}\frac{N_1^{\alpha-1}}{N_1^{\alpha-1}+N_2^{\alpha-1}}\right) M .
\end{align}
Now, we consider the case $N_1 \ge N_2 > 1$.
Then, noting that
\begin{align}
	\nonumber
	& 1-\frac{1}{N_1}\frac{N_1^{\alpha-1}}{N_1^{\alpha-1}+N_2^{\alpha-1}} \simeq 1 \\
	\nonumber
	& 1-\frac{1}{N_2}\frac{N_1^{\alpha-1}}{N_1^{\alpha-1}+N_2^{\alpha-1}} \simeq 1 ,
\end{align}
we get
\begin{align}
	\nonumber
	\frac{u_{1j(\mathrm{DD})}}{u_{2j(\mathrm{DD})}} &= \left(\frac{N_1}{N_2}\right)^{\alpha-2} \\
	\nonumber
	& \left\{ \begin{array}{ll}
		\ge 1 & (\alpha > 2) \\
		< 1 & (\alpha < 2) . \\
	\end{array} \right.
\end{align}

Third, for $1 < \alpha < 1.5$ (see Fig.~S4),
\begin{align}
	\nonumber
	& u_{1j(\mathrm{CC})} > u_{2j(\mathrm{CC})} \\
	\nonumber
	& u_{1j(\mathrm{CD})} > u_{2j(\mathrm{CD})} \\
	\nonumber
	& u_{1j(\mathrm{DC})} < u_{2j(\mathrm{DC})} \\
	\nonumber
	& u_{1j(\mathrm{DD})} < u_{2j(\mathrm{DD})} .
\end{align}

\begin{figure}[H]
\begin{center}
\includegraphics[width=0.5\textwidth]{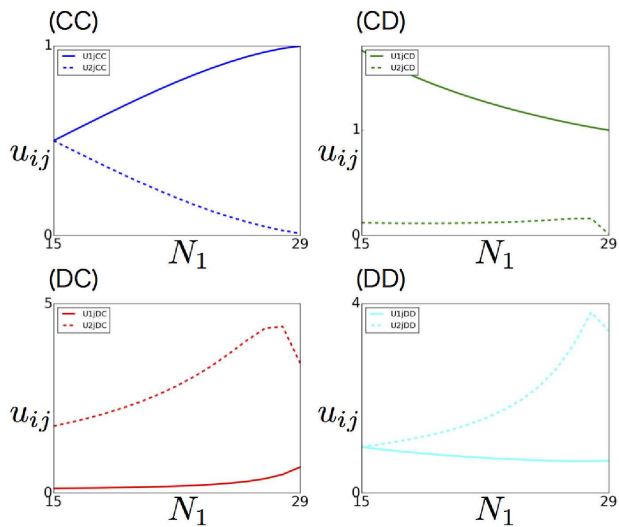}
\caption{Individual payoff $u_{1j(\mathrm{YZ})}$ (solid line) and $u_{2j(\mathrm{YZ})}$ (broken line) plotted as a function of $N_1$ for $N=30, \alpha=1.2$.
The blue (upper left), green (upper right), red (lower left), and cyan (lower right) lines indicate $\mathrm{YZ}=\mathrm{CC}, \mathrm{CD}, \mathrm{DC}$, and $\mathrm{DD}$, respectively.
}
\end{center}
\end{figure}
The difference between $1<\alpha<1.5$ and $1.5<\alpha<2$ lies in the behaviors of $u_{ij(\mathrm{DC})}$. For $1<\alpha<1.5$, individuals in the smaller group always receive higher payoffs than those in the larger group do in the case of $\mathrm{DC}$. Here, $u_{1j(\mathrm{DC})}$ and $u_{2j(\mathrm{DC})}$ are given by
\begin{align}
	\nonumber
	& u_{1j(\mathrm{DC})} = \frac{1}{N_1}\frac{N_1^{\alpha-1}}{N_1^{\alpha-1}+N_2^{\alpha}} \left(1-\frac{1}{N_1}\frac{N_2^{\alpha}}{N_1^{\alpha-1}+N_2^{\alpha}}\right) M \\
	\nonumber
	& u_{2j(\mathrm{DC})} = \frac{1}{N_2}\left(\frac{N_2^{\alpha}}{N_1^{\alpha-1}+N_2^{\alpha}}\right)^2 M .
\end{align}
Recall that $u_{1j(\mathrm{DC})} > u_{2j(\mathrm{DC})}$ holds for $N_1 \gg N_2, 1.5 < \alpha < 2$. Thus, we consider the transition for $N_1 \gg N_2 > 1$, where we get
\begin{align}
	\nonumber
	\frac{u_{1j(\mathrm{DC})}}{u_{2j(\mathrm{DC})}} & = \frac{N_1^{\alpha-3}(N_1^{\alpha}+(N_1-1)N_2^{\alpha})}{N_2^{2\alpha-1}} \\
	\nonumber
	& \simeq N_1^{2\alpha-3}\frac{N_2^{\alpha}+1}{N_2^{2\alpha-1}} \\
	\nonumber
	& \left\{ \begin{array}{ll}
		\gtrsim 1 & (\alpha > 1.5) \\
		\lesssim 1 & (\alpha < 1.5) . \\
	\end{array} \right.
\end{align}
Then, $u_{1j(\mathrm{DC})} < u_{2j(\mathrm{DC})}$ always holds in the case of $\alpha<1.5$.

Fourth, for $0.5 < \alpha < 1$ (see Fig.~S5),
\begin{align}
	\nonumber
	& u_{1j(\mathrm{CC})} > u_{2j(\mathrm{CC})} \\
	\nonumber
	& u_{1j(\mathrm{CD})} \left\{ \begin{array}{ll}
		> u_{2j(\mathrm{CD})} & (N_1 \simeq N_2) \\
		< u_{2j(\mathrm{CD})} & (N_1 \gg N_2) \\
	\end{array} \right. \\
	\nonumber
	& u_{1j(\mathrm{DC})} < u_{2j(\mathrm{DC})} \\
	\nonumber
	& u_{1j(\mathrm{DD})} < u_{2j(\mathrm{DD})} .
\end{align}

\begin{figure}[H]
\begin{center}
\includegraphics[width=0.5\textwidth]{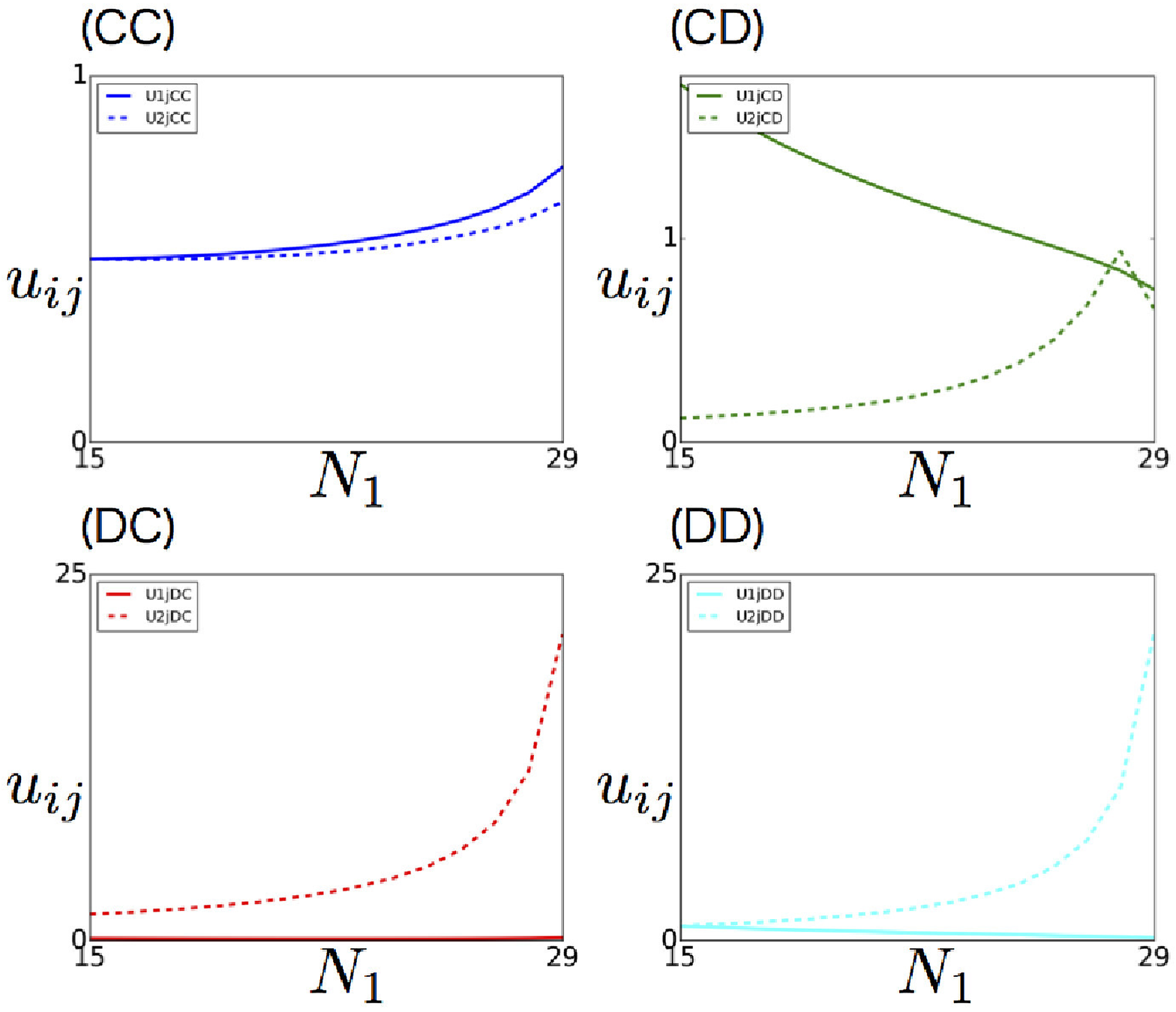}
\caption{Individual payoff $u_{1j(\mathrm{YZ})}$ (solid line) and $u_{2j(\mathrm{YZ})}$ (broken line) plotted as a function of $N_1$ for $N=30, \alpha=0.52$.
The blue (upper left), green (upper right), red (lower left), and cyan (lower right) lines indicate $\mathrm{YZ}=\mathrm{CC}, \mathrm{CD}, \mathrm{DC}$, and $\mathrm{DD}$, respectively.
}
\end{center}
\end{figure}
The difference between $0.5 < \alpha < 1$ and $1 < \alpha < 1.5$ lies in the behavior of $u_{ij(\mathrm{CD})}$. For $0.5 < \alpha < 1$, individuals in the smaller group receive higher payoffs than those in the larger one do in the case of $\mathrm{CD}$ for $N_1 \gg N_2$
\footnote{As an exception, for $N_2=1$, $u_{1j(\mathrm{CD})} > u_{2j(\mathrm{CD})}$ holds. When $\alpha$ is much smaller ($\alpha<0.5$), $u_{1j(\mathrm{CD})} < u_{2j(\mathrm{CD})}$ holds even for $N_2=1$.}

Here, $u_{1j(\mathrm{CD})}$ and $u_{2j(\mathrm{CD})}$ are given by
\begin{align}
	\nonumber
	& u_{1j(\mathrm{CD})} = \frac{1}{N_1}\left(\frac{N_1^{\alpha}}{N_1^{\alpha}+N_2^{\alpha-1}}\right)^2 M \\
	\nonumber
	& u_{2j(\mathrm{CD})} = \frac{1}{N_2}\frac{N_1^{\alpha-1}}{N_1^{\alpha}+N_2^{\alpha-1}} \left(1-\frac{1}{N_2}\frac{N_1^{\alpha}}{N_1^{\alpha}+N_2^{\alpha-1}}\right) M .
\end{align}
Then, assuming $N_1 \gg N_2 > 1$, we obtain
\begin{align}
	\nonumber
	\frac{u_{2j(\mathrm{CD})}}{u_{1j(\mathrm{CD})}} & = \frac{N_2^{\alpha-3}((N_2-1)N_1^{\alpha}+N_2^{\alpha})}{N_1^{2\alpha-1}} \\
	\nonumber
	& \simeq N_1^{1-\alpha}N_2^{\alpha-3}(N_2-1) \\
	\nonumber
	& \left\{ \begin{array}{ll}
		\lesssim 1 & (\alpha > 1) \\
		\gtrsim 1 & (\alpha < 1) \\
	\end{array} \right.
\end{align}
Thus, $u_{1j(\mathrm{CD})} < u_{2j(\mathrm{CD})}$ sometimes holds in the case of $0.5 < \alpha < 1$.

Fifth, for $\alpha < 0.5$ (see Fig.~S6),
\begin{align}
	\nonumber
	& u_{1j(\mathrm{CC})} < u_{2j(\mathrm{CC})} \\
	\nonumber
	& u_{1j(\mathrm{CD})} \left\{ \begin{array}{ll}
		> u_{2j(\mathrm{CD})} & (N_1 \simeq N_2) \\
		< u_{2j(\mathrm{CD})} & (N_1 \gg N_2) \\
	\end{array} \right. \\
	\nonumber
	& u_{1j(\mathrm{DC})} < u_{2j(\mathrm{DC})} \\
	\nonumber
	& u_{1j(\mathrm{DD})} < u_{2j(\mathrm{DD})} .
\end{align}

\begin{figure}[H]
\begin{center}
\includegraphics[width=0.5\textwidth]{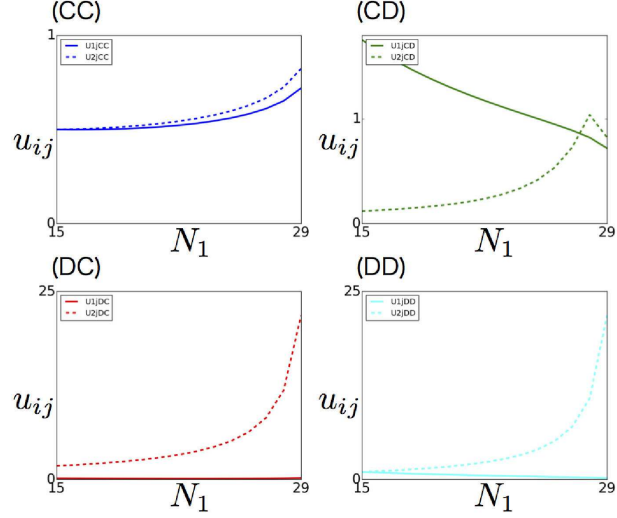}
\caption{Individual payoff $u_{1j(\mathrm{YZ})}$ (solid line) and $u_{2j(\mathrm{YZ})}$ (broken line) plotted as a function of $N_1$ for $N=30, \alpha=0.48$. The blue (upper left), green (upper right), red (lower left), and cyan (lower right) lines indicates $\mathrm{YZ}=\mathrm{CC}, \mathrm{CD}, \mathrm{DC}$, and $\mathrm{DD}$, respectively.
}
\end{center}
\end{figure}
The difference between $\alpha < 0.5$ and $0.5 < \alpha < 1$ lies in the behavior of $u_{ij(\mathrm{CC})}$. For $0.5 < \alpha < 1$, individuals in the smaller group always receive higher payoffs than those in the larger one do the case of $\mathrm{CC}$.
Here, $u_{1j(\mathrm{CC})}$ and $u_{2j(\mathrm{CC})}$ are given by
\begin{align}
	\nonumber
	& u_{1j(\mathrm{CC})} = \frac{1}{N_1}\left(\frac{N_1^{\alpha}}{N_1^{\alpha}+N_2^{\alpha}}\right)^2 M \\
	\nonumber
	& u_{2j(\mathrm{CC})} = \frac{1}{N_2}\left(\frac{N_2^{\alpha}}{N_1^{\alpha}+N_2^{\alpha}}\right)^2 M .
\end{align}
Then, we obtain
\begin{align}
	\nonumber
	\frac{u_{2j(\mathrm{CC})}}{u_{1j(\mathrm{CC})}} & = \left(\frac{N_2}{N_1}\right)^{2\alpha-1} \\
	\nonumber
	& \left\{ \begin{array}{ll}
		< 1 & (\alpha > 0.5) \\
		> 1 & (\alpha < 0.5) . \\
	\end{array} \right.
\end{align}
Thus, $u_{1j(\mathrm{CC})} < u_{2j(\mathrm{CC})}$ always holds in the case of $\alpha < 0.5$.

\end{document}